\documentclass[10 pt, conference]{ieeeconf}
\IEEEoverridecommandlockouts
\usepackage{cite}
\usepackage{amsthm}
\usepackage{amsmath,amssymb,amsfonts}
\usepackage{algorithmic,algorithm} 
\usepackage{graphicx}
\usepackage{textcomp}
\usepackage{amssymb}
\usepackage{xcolor}
\usepackage{bm}

\usepackage[T1]{fontenc}
\usepackage[colorinlistoftodos]{todonotes}
\usepackage{makecell}
\def\BibTeX{{\rm B\kern-.05em{\sc i\kern-.025em b}\kern-.08em
		T\kern-.1667em\lower.7ex\hbox{E}\kern-.125emX}}

\usepackage{multirow}
\usepackage{mathrsfs}
\DeclareMathAlphabet{\mathcal}{OMS}{cmsy}{m}{n}
\usepackage{pgfplots}
\usepackage{tikz}
\usetikzlibrary{decorations.pathreplacing,calligraphy}
\usetikzlibrary{arrows}
\usetikzlibrary{arrows.meta}

\newcommand{\Eb}{{\mathbb{E}}}
\newcommand{\Rb}{{\mathbb{R}}}
\newcommand{\Cb}{{\mathbb{C}}}

\newcommand{\Zb}{{\mathbb{Z}}}

\newcommand{\Ob}{{\mathbb{O}}}

\newcommand{\Vc}{{\mathcal{V}}}

\newcommand{\Fc}{{\mathcal{F}}}

\newcommand{\Qc}{{\mathcal{Q}}} 

\newcommand{\Cc}{{\mathcal{C}}}
\newcommand{\Ic}{{\mathcal{I}}}

\newcommand{\Tc}{{\mathcal{T}}}
\newcommand{\Sc}{{\mathcal{S}}}

\newcommand{\Jc}{{\mathcal{J}}}

\newcommand{\tAi}{\tilde{A}_i}
\newcommand{\tCi}{\tilde{C}_i}

\newcommand{\ra}{{\rightarrow}}

\newcommand{\rs}{\text{rowspan}}

\newcommand{\rd}{{\mathrm{d}}}

\DeclareMathOperator{\cov}{Cov}

\DeclareMathOperator{\spe}{sp}

\DeclareMathOperator{\Span}{{span}}
\DeclareMathOperator{\med}{{med}}

\theoremstyle{plain}

\newtheorem{lemma}{\textbf{Lemma}}
\newtheorem{theorem}{\textbf{Theorem}}
\newtheorem{remark}{\textbf{Remark}}
\newtheorem{assumption}{\textbf{Assumption}}

\newtheorem{definition}{\textbf{Definition}}
\newtheorem{problem}{\textbf{Problem}}

\title{Secure State Estimation with Asynchronous Measurements against Malicious  Measurement-data and Time-stamp Manipulation\\
	\thanks{Identify applicable funding agency here. If none, delete this.}
}
\author{Zishuo Li, Anh Tung Nguyen, André Teixeira, Yilin Mo, Karl H. Johansson}
\begin{document}
	
	\maketitle
	
	\begin{abstract}
		This paper proposes a secure state estimation scheme with non-periodic asynchronous measurements for linear continuous-time systems under false data attacks on the measurement transmit channel. 
		After sampling the output of the system, a sensor transmits the measurement information in a triple composed of sensor index, time-stamp, and measurement value to the fusion center via vulnerable communication channels. 
		The malicious attacker can corrupt a subset of the sensors through (i) manipulating the time-stamp and measurement value; (ii) blocking transmitted measurement triples; or (iii) injecting fake measurement triples. 
		To deal with such attacks,
		we propose the design of local estimators based on observability space decomposition, where each local estimator updates the local state and sends it to the fusion center after sampling a measurement.
		Whenever there is a local update, the fusion center combines all the local states and generates a secure state estimate by adopting the median operator.
		We prove that local estimators of benign sensors are unbiased with stable covariance. 
		Moreover, the fused central estimation error has bounded expectation and covariance against at most $p$ corrupted sensors as long as the system is $2p$-sparse observable.
		The efficacy of the proposed scheme is demonstrated through an application on a benchmark example of the IEEE 14-bus system.

		\end{abstract}
		
		\section{Introduction}
		Many real-world large-scale systems, such as power systems, water distribution networks, and transportation networks, are examples of cyber-physical systems where physical processes are tightly coupled with digital devices and infrastructure. 
		Due to their large-scale description, those cyber-physical systems are generally divided into many smaller partitions that are jointly monitored and controlled via wired or wireless communications, leaving the systems vulnerable to malicious attackers.
		The attackers might exploit unprotected communication channels to manipulate data shared among the systems, with the purpose of intentionally disrupting the systems. 
		For example, reports have shown the disastrous consequences of malicious software called Stuxnet on industrial control systems in Iran \cite{falliere2011w32}.
		In 2016, massive damage caused by Sandworm was witnessed in a Ukrainian power grid \cite{kshetri2017hacking}.
		Motivated by these and many other examples in \cite{kshetri2017hacking}, security in cyber-physical systems has received much attention from the control system society. 
		In particular, the challenge of securely estimating unmeasured states under threats has been widely addressed \cite{liu2020local,li2023efficient,fawzi2014secure,shoukry2015event,nakahira2018attack}, given the crucial role of state estimation in control systems.
		This challenge, known as secure state estimation of control systems, is mainly addressed throughout this paper.
		\begin{figure}[!ht]
			\centering
			\includegraphics[width=\linewidth]{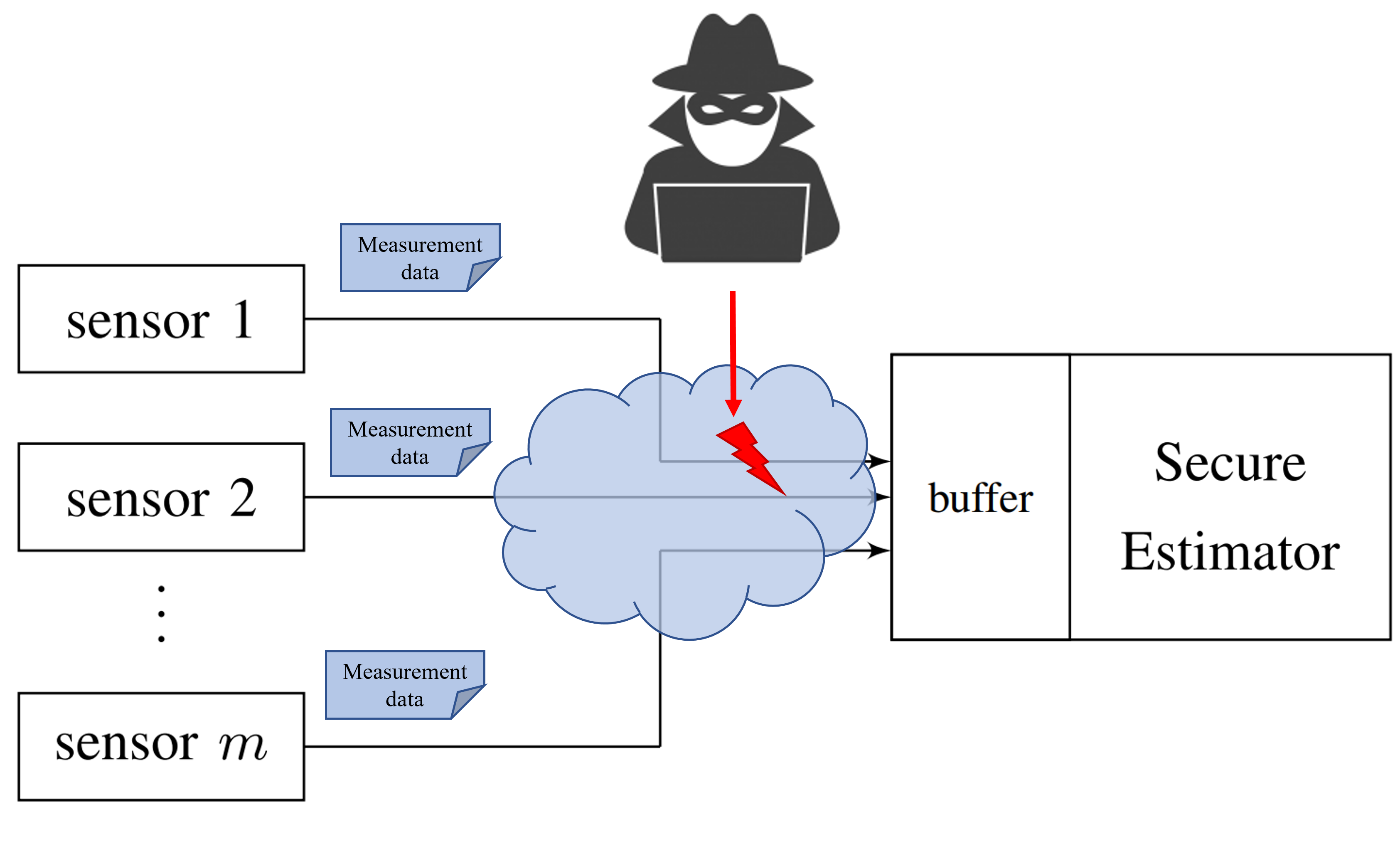}
			\caption{
				The general state estimation scheme with a measurement buffer dealing with out-of-order sequence \cite{Kaempchen2003DATASS,out-of-sequence-measurements-IVS,FIR_Discretely_Delayed} where some of the communication channels are compromised by a malicious adversary.} \label{fig:flow}
    \vspace{-18pt}
		\end{figure}
		
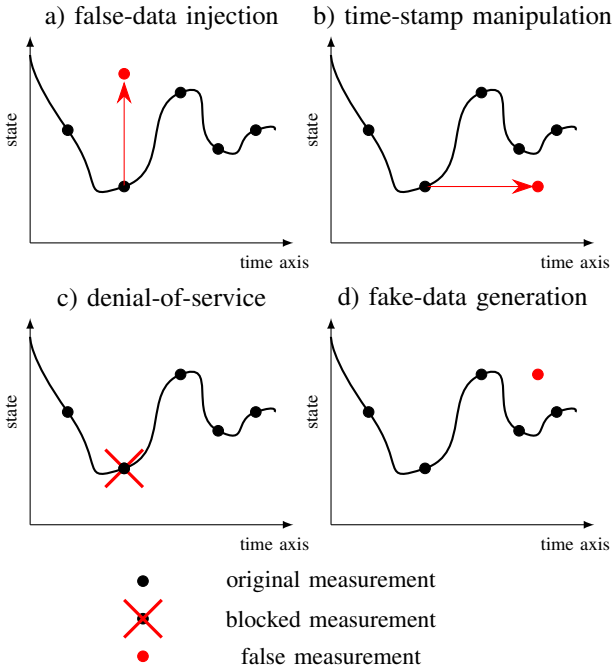
\begin{figure}[!ht]
			\centering
			\begin{tikzpicture}
%
%
%
%
%
%

\draw[-latex][black] (0, 0)--(3.5,0);
\draw[-latex][black] (0, 0)--(0,2.75);
\node [rotate=90] at (-0.25,1.5) {\scriptsize state};
\node at (3.25,-0.25) {\scriptsize time axis};

\draw[-latex][black] (4,0)--(7.5,0);
\draw[-latex][black] (4,0)--(4,2.75);
\node [rotate=90] at (3.75,1.5) {\scriptsize state};
\node at (7.25,-0.25) {\scriptsize time axis};

\draw[-latex][black] (0,-3.75)--(3.5,-3.75);
\draw[-latex][black] (0,-3.75)--(0,-1);
\node [rotate=90] at (-0.25,-2.25) {\scriptsize state};
\node at (3.25,-4) {\scriptsize time axis};

\draw[-latex][black] (4,-3.75)--(7.5,-3.75);
\draw[-latex][black] (4,-3.75)--(4,-1);
\node [rotate=90] at (3.75,-2.25) {\scriptsize state};
\node at (7.25,-4) {\scriptsize time axis};

\filldraw (1.25,0.75) circle (2pt);
\filldraw [red] (1.25,2.25) circle (2pt);

\filldraw (0.5,1.5)  circle (2pt);
\filldraw (2,2)  circle (2pt);
\filldraw (2.5,1.25)  circle (2pt);
\filldraw (3,1.5)   circle (2pt);

   \draw[-{Stealth[length=3mm, width=2mm]}][red](1.25,0.75)--(1.25,2.15);
\node at (1.75,3) {a) false-data injection};

\draw[thick] plot [smooth,tension=1.5] coordinates{(0,2.5) (0.5,1.5) (1.25,0.75) (2,2) (2.5,1.25) (3,1.5) (3.25,1.5)};

\filldraw (5.25,0.75) circle (2pt);
\filldraw [red] (6.75,0.75) circle (2pt);

\filldraw (4.5,1.5)  circle (2pt);
\filldraw (6,2)  circle (2pt);
\filldraw (6.5,1.25)  circle (2pt);
\filldraw (7,1.5)   circle (2pt);

   \draw[-{Stealth[length=3mm, width=2mm]}][red] (5.25,0.75)--(6.7,0.75);
\node at (5.75,3) {b) time-stamp manipulation};
\draw[thick] plot [smooth,tension=1.5] coordinates{(4,2.5) (4.5,1.5) (5.25,0.75) (6,2) (6.5,1.25) (7,1.5)  (7.25,1.5)};

\filldraw (1.25,-3) circle (2pt);

\filldraw (0.5,-2.25)  circle (2pt);
\filldraw (2,-1.75)  circle (2pt);
\filldraw (2.5,-2.5)  circle (2pt);
\filldraw (3,-2.25)   circle (2pt);

\node at (1.75,-0.75) {c) denial-of-service};
\draw [red][line width=1.2pt](1,-3.25)--(1.5,-2.75);
\draw [red][line width=1.2pt](1.5,-3.25)--(1,-2.75);

\draw[thick] plot [smooth,tension=1.5] coordinates{(0,-1.25) (0.5,-2.25) (1.25,-3) (2,-1.75) (2.5,-2.5) (3,-2.25)  (3.25,-2.25)};

\filldraw (1.25,-3) circle (2pt);
\filldraw [red] (6.75,-1.75) circle (2pt);

\filldraw (4.5,-2.25)  circle (2pt);
\filldraw (5.25,-3)  circle (2pt);
\filldraw (6,-1.75)  circle (2pt);
\filldraw (6.5,-2.5)  circle (2pt);
\filldraw (7,-2.25)   circle (2pt);

\node at (5.75,-0.75) {d) fake-data generation};

\draw[thick] plot [smooth,tension=1.5] coordinates{(4,-1.25) (4.5,-2.25) (5.25,-3) (6,-1.75) (6.5,-2.5) (7,-2.25)  (7.25,-2.25)};


\filldraw (1.5,-4.5)  circle (2pt); 
\node at (4.,-4.5) {\small original measurement};

\filldraw (1.5,-5)  circle (2pt); 
\draw [red][line width=1.2pt](1.25,-4.75)--(1.75,-5.25);
\draw [red][line width=1.2pt](1.75,-4.75)--(1.25,-5.25);
\node at (4.,-5) {\small blocked measurement};

\filldraw [red] (1.5,-5.5)  circle (2pt); 
\node at (4.,-5.5) {\small false measurement};

\end{tikzpicture}
			\caption{Examples of  the spatio-temporal false data attack that can manipulate both the time-stamp value and the measurement value.} \label{fig:time_data_attack}
   \vspace{-18pt}
		\end{figure}

		
        In asynchronous sampled systems where data is sampled at different rates, the time-stamp of the measurement, sent to the estimator, is vulnerable to malicious adversaries.
		In this paper, we propose a novel model of false data attack (see Fig \ref{fig:time_data_attack}) on the asynchronous non-periodic sampled system which includes both integrity attacks such as false-data injection \cite{hu_stateestimation_automatica}, and availability attacks such as denial-of-service attacks \cite{lu2019resilient,su2018cooperative,liu2021resilient}. Moreover, we investigate the influence of time-stamp information in the model and include the time-stamp manipulation attack. 
  {Denial-of-service attacks might restrict the availability of the measurement by blocking the measurement data transmitted to the estimator.}
  Besides that, for the asynchronous non-periodic sampled system, injecting fake data into authentic measurement streams is not easily detected but is harmful to the system. 
  We also include this fake data generation attack in our model, and there may be combinations of those four types of attacks.
		
		\textbf{Related works:}

        To deal with the problem of secure state estimation against false data injection attacks, three research directions consisting of the sliding window method, the estimator switching method, and the local decomposition-fusion method, have been mainly developed in recent years.
        The sliding window method considers the past sensor measurements in a finite time horizon to do the state estimation via batch optimization problems. 
        With the help of certain observability redundancy conditions, the authors  in \cite{fawzi2014secure,shoukry2015event} deal with the minimization problem of the norm of residual signal to obtain the secured state estimation.
        The estimator switching method maintains multiple parallel estimators that utilize the measurements given by a subset of all the sensors.
	To obtain the secure station estimation, a detection algorithm is proposed to choose the trusted subsets of sensors whose corresponding local estimates are then fused \cite{Mishra2017TCNS,nakahira2018attack}.
	In the local decomposition fusion, multiple decentralized estimators, each of which samples the measurement of one local sensor, are designed to solve a combinatorial complex problem of the full state estimation. 
		Then, local estimated states  provided by such local state estimators are fused by a convex optimization problem designed to generate secure estimation \cite{liu2020local,li2023efficient}.

Denial-of-service attacks block the measurement transmitted to the state estimator, worsening the state estimation performance. To handle such attacks conducted to multiple transmission channels, a class of partial observers that provide reliable partial state estimates is proposed in \cite{lu2019resilient}. 
The authors in \cite{su2018cooperative} propose a detection-compensation scheme to detect the presence of DoS attacks and then effectively reconstruct missing state estimates through past available states. This scheme eventually  mitigates the attack impact on the performance of the state estimation.
On the other hand, event-triggered mechanisms have received much attention since it deals with DoS attacks very effectively. Flexible event-triggered mechanisms such as a newly proposed dynamic event-triggered state estimation \cite{liu2021resilient} are very efficient in alleviating the performance loss of the state estimation caused by DoS attacks. 



    To the best of our knowledge, comparatively little progress has been made toward studying the influence of time-stamp manipulation on estimation performance.
    Li et al. \cite{LI2008199} and Guo et al. \cite{guo_multi_async} propose Kalman filter-based algorithms for non-uniformly sampled multi-rate systems. To deal with the problem of the asynchronous sampling linear and nonlinear systems, the authors in
    \cite{Feddaoui_asy_kalman} propose a class of continuous-discrete observers, resulting in a differential  Riccati equation. Moreover, the stability of the corresponding differential Riccati equation is proved, which guarantees the convergence of the observer. 
    Ding et al.\cite{DING2009324} and Muhammad et al.\cite{Muhammad_pathological_sample} analyze the observability degradation problem of multi-rate sampling and non-periodic sampling systems.
    For time-stamp-related attacks, the impact of Time Synchronization Attack (TAS) on smart grids is studied by the authors in  \cite{zhang_TSA_TSG}. 

		In this paper, we first propose a novel spatio-temporal false data attack model. Then, we design a secure estimation algorithm to recover the system state in the presence of such a spatio-temporal false data attack on {fixed $p$ sensors}. The algorithm has the following merits:
        \begin{enumerate}
            \item  The algorithm adopts an asynchronous non-periodic sampling framework and thus is general enough to include synchronous sampling and multi-rate sampling scenarios.
            \item The negative impact of the spatio-temporal false data attack is transformed into the value change of local state estimations. This transformation allows us to handle such an attack by using a resilient fusion algorithm that can generate secure estimates.
            \item 
            {Given the $2p$-sparse observability of the system, we show that the proposed estimation has stable error.}
            Moreover, we give explicit estimation error expectation and covariance bounds in the presence of such an attack.
        \end{enumerate}

		\textit{Notations:}
		The sets of {positive integers, non-negative integers, and non-negative real numbers are denoted as $\Zb_{>0},\Zb_{\geq0}$, and $\Rb_{\geq0}$, respectively. }
		The cardinality of a set $\Sc$ is denoted as $|\Sc|$.  Denote the span of row vectors of matrix $A$ as $\rs(A)$. 
		All-zero matrices with an appropriate size are denoted as $\mathbf{0}$. 
		We denote $I$ as an identity matrix with an appropriate dimension. Define $\mathbf{e}_j$ as a canonical basis vector with the appropriate dimension where $1$ is on the $j$-th entry and $0$ is on all the other entries.
		The spectral radius of matrix $A$ is denoted as $\rho(A)$.
		For a vector $x$, $[x]_j$ stands for its $j$-th entry.
		We denote the continuous time index in a pair of parenthesis $(\cdot)$ and the discrete-time index in a pair of brackets $[\cdot]$.
		For notation simplicity, we denote time index $(t_k)$ as $[k]$ for continuous variables. 
		
		\section{Problem Formulation and Preliminaries}
		\label{sec:problem}	
		In this section, we first introduce the system, the modeling of asynchronous measurements, and several assumptions that will be used throughout this paper. Secondly, we present a novel spatio-temporal false data attack. Finally, the secure estimation problem we intend to solve is formulated.
		\subsection{State Estimation with Asynchronous Measurements}\label{sec:basic_measure}
		We consider a continuous-time linear-invariant system:
		\begin{align}\label{eq:system}
			\dot{x}(t)=A x(t)+w(t)
		\end{align}
		where $x(t) \in \mathbb{R}^{n}$ is the system state and the process noise $w(t)$ is a Wiener process. The process noise  from time $t_1$ to $t_2$ is denoted as $w(t_1,t_2)$ and the corresponding covariance is $Q\cdot (t_2-t_1)$.
		The initial state
		$x(0)$ is assumed to be a Gaussian random vector with a known expectation and is independent of the measurement noise.
		We introduce the following assumption on $A$ to prevent system observability degradation problems.   
		\begin{assumption}\label{as:geo_mul}
			The geometric multiplicity of all the eigenvalues of $A$ is 1. 
		\end{assumption}
		Since one can perform similarity transformations on the system states and study the transformed system where $A$ is in the Jordan canonical form, without loss of generality, we assume that $A$ is in the Jordan canonical form throughout this paper.
		

		Let us denote the sensor index set as $\Ic \triangleq\{1,2, \ldots, m\}$ and the state index set as $\Jc \triangleq\{1,2, \ldots, n\}$. We consider a general asynchronous non-periodic sampled scenario where the estimation operator receives measurement triples from sensor $i \in \Ic$ of the form:
		\begin{align*}
			\textbf{measurement triple: }(i, t, y_i(t)),
		\end{align*}
		where $i$ is the sensor index, $t$ is the time-stamp, and $y_i(t)$ is the measurement value given by sensor $i$. 
		We consider scalar measurement values with measurement model as
		\begin{equation}\label{eq:y_i_def}
			y_i(t)=C_i x(t)+v_i(t), 
		\end{equation} 
		where $C_i\in\Rb^{1\times n}$ is the measurement matrix and $v_i(t) \in \mathbb{R}$  is Gaussian measurement noise with time-varying covariance $R_i(t)$ 
		which is under the following assumption.
		\begin{assumption}\label{as:R}
			For every sensor $i \in \Ic$, its corresponding measurement noise covariance $R_i(t)$
			satisfies the following:
			\begin{align*}
				0\leq R_i(t) \leq \bar{r},\quad  \forall t\in\Rb_{\geq0},
			\end{align*}
			where $\bar{r}$ is a given positive constant scalar.
		\end{assumption}
		We need observability redundancy to obtain secure estimate.
		\begin{assumption}\label{as:sparse_obs}
			The system $(A,C)$ is {$2p$-sparse observable}, i.e., the system $(A,C_{\Ic\setminus\Cc})$ is observable\footnote{The matrix $C_{\Ic\setminus\Cc}$ represents the matrix composed of rows of $C$ with row index in $\Ic\setminus\Cc$.} for any subset of sensors $\Cc\subset\Ic$ with cardinality $|\Cc| = 2p$.  
		\end{assumption}

		Define the set of sampling time-stamps from sensor $i$ as $\Gamma_i$. The set of all sampling time-stamps from all the sensors is denoted as $\Gamma\triangleq \bigcup_{i=1}^m \Gamma_i$. Without loss of generality, the time when the estimation starts to work is set as $t_0=0$.

		In order to guarantee system observability under non-uniform asynchronous measurements, we introduce the following notations and assumptions.
        Define the set of sampling time intervals and cumulative sampling time from sensor $i$ as follows
		\begin{align*}
			\Tc_i &\triangleq \left\{t_k-t_{k-1}\ | ~t_k,t_{k-1}\in\Gamma_i, k\in\Zb_{>0} \right\},\
			\Tc\triangleq \bigcup_{i=1}^m \Tc_i,\\
            \widetilde{\Tc}_i &\triangleq \left\{t_k-t_{0}\ | ~t_k,\in\Gamma_i, k\in\Zb_{>0} \right\},\
			\widetilde{\Tc}\triangleq \bigcup_{i=1}^m \Tc_i.
		\end{align*}
        Define the system pathological sampling interval set~\cite{DING2009324} as
        \begin{align*}
            \Tc^*\triangleq \left\{ T>0 | \exp(\lambda_i T)=\exp(\lambda_j T),  i\neq j, \right.\\
            ~~~~~
            \left.\lambda_i,\lambda_j\in \spe(A) \subseteq \Cb \right\} .
		\end{align*} 
		To prevent system observability degradation problems due to discrete-time samplings, the following assumption, which is also seen in \cite{Muhammad_pathological_sample,DING2009324}, is introduced.
		\begin{assumption}[Non-pathological sampling time]\label{as:sample_time}
			The sampling time interval set $\Tc$ satisfies the following conditions:
			\begin{align*}
				\sup \Tc \leq T_{\max},~ \text{and}~ \widetilde{\Tc}\cap \Tc^*=\varnothing .
			\end{align*}
		\end{assumption} 
		
		\begin{remark}
			Since Assumption \ref{as:geo_mul} ensures that $\lambda_i\neq\lambda_j$ where $i\neq j$,
			the system pathological sampling interval set $\Tc^*$ is equivalent to the following set:
			\begin{align*}
				\left\{ \left.\frac{2k\pi \sqrt{-1}}{\lambda_i-\lambda_j}\right| \lambda_i,\lambda_j\in\spe(A)\subset \Cb,i\neq j,  k\in\Zb\right\}\bigcap \Rb_{> 0} .
			\end{align*}
			In the designing phase, the set $\Tc^*$ can be calculated given matrix $A$ in \eqref{eq:system}. Then, measurement triples that contain sampling time intervals in $\Tc^*$ or larger than $T_{\max}$ are discarded.
			Thus, with minor information loss, Assumption~\ref{as:sample_time} always holds. 
		\end{remark}
		
		
		\subsection{Measurement-data and time-stamp manipulation} 
		In this section, we introduce a general type of cyber-attacks called a spatio-temporal false data attack that generalizes the integrity attacks and the availability attacks~ (see Fig.~\ref{fig:time_data_attack} for more detail).
		For the convenience of denotation about the measurement sampling process, we introduce the measurement triple generation set as all the measurement triples with time-stamp $t$: 
		\begin{align*}
			\Sc(t)= \left\{(i,t,y_i(t), i \in \Ic \right\}. 
		\end{align*}
		Moreover, $\Sc^a(t)$ denotes the set of manipulated measurement triples with time-stamp $t$.
		The spatio-temporal false data attack is defined as follows:
		\begin{definition}[spatio-temporal false data attack]\label{def:attack}
			The attacker can manipulate measurement triples given by corrupted sensor $i\in\Cc$ in the following four ways:
            \begin{align*}
                \textbf{(i) false-data injection } & ( i, t,y^a_i(t))\leftarrow ( i, t,y_i(t) ),\\
                \textbf{(ii) time-stamp manipulation } & ( i, t^a,y_i(t))\leftarrow( i, t,y_i(t) ),
            \end{align*}
			where $( i,t,y_i(t))\in \Sc(t)$. $y^a_i(t)$ and $t^a$ denotes the manipulated data.
            \begin{align*}
                \textbf{(iii) denial-of-service } & \Sc^a(t)\leftarrow \Sc(t)\setminus ( i,t,y_i(t)),\\
                \textbf{(iv) fake-data generation } & \Sc^a(t)\leftarrow \Sc(t)\cup \left( i,t^f,y^f_i(t)\right),
            \end{align*}
            where $( i,t,y_i(t))\in \Sc(t)$, $( i,t^f,y^f_i(t))\notin \Sc(t)$, the superscript ``$f$" means ``fake" (not real measurment).
			
			If the set of corrupted sensors (denoted by $\Cc$) satisfies $|\Cc|\leq p$, we say that it is an $p$-sparse spatio-temporal false data attack. 	
		\end{definition}

		In the scope of this paper, we study the $p$-sparse spatio-temporal false data attack.
		The estimation scheme has no direct information about the set of corrupted sensors $\Cc$, but it knows the maximum number of corrupted sensors denoted by $p$.
		The manipulated time-stamp set $\Gamma^a$ and manipulated time intervals $\Tc^a$ are defined as follows:
		\begin{align}
			\Gamma^a\triangleq &\bigcup_{i=1}^m \Gamma^a_i,&\Gamma^a_i\triangleq \{t~|(i,t,y_i(t))\in\Sc^a(t)\}.
		\end{align}

		Due to various delays, received measurement time-stamps may not be in increasing order, resulting in the out-of-sequence problem \cite{Kaempchen2003DATASS,out-of-sequence-measurements-IVS,FIR_Discretely_Delayed}. 
		This problem is generally dealt with by utilizing the \emph{Buffering} method.
		The buffer simply stores all the measurements from a time window of length $d$ before sending them to the fusion center, where $d$ is the maximum delay of a measurement sample \cite{Kaempchen2003DATASS,out-of-sequence-measurements-IVS,FIR_Discretely_Delayed}.
		Measurements delayed more than $d$ are seen as non-informative and discarded from the buffer. 
		In this way, the measurement sequence after the buffer is sorted in the correct order. 
		In this paper, we assume a similar buffering system is working before the secure estimator (see Fig.~\ref{fig:flow}),  yielding the following assumption.
		\begin{assumption}\label{as:time_in_order}
			The incoming manipulated time-stamp is in a strictly increasing order, i.e., $\Gamma^a=\{t_0,t_1,t_2,\cdots\}$ and $0 = t_0 < t_i < t_{i+1},~(\forall i \in\Zb_{>0})$.
		\end{assumption}
		\subsection{Secure estimation problem}

		The communication protocol among sensors and the measurement buffer depicted in Fig.~\ref{fig:flow} leaves the system vulnerable to spatio-temporal false data attacks in Definition~\ref{def:attack} which includes measurement-data and time-stamp manipulations.
		To estimate the system states under such attacks, this paper will propose a solution to deal with a secure state estimation problem, which is defined below.
		\begin{problem}[Secure state estimation]\label{pb:secure_est}
			Find an estimator which is a measurable time-varying function $f_{t}(\cdot)$  of all manipulated history measurement triples:
			$$\hat{x}(t)=f_{t}(\Sc^a(\tau),\tau\leq t)$$ 
			such that the estimation error expectation and covariance are uniformly bounded at sampling instants:
			\begin{align*}		
				\sup_{t\in\Gamma^a}&~\|\Eb\left[\hat{x}(t)-         x(t)\right]\|_{\infty} \leq \gamma_{e}(A,C,Q,R) ,\\
				\sup_{t\in\Gamma^a}&~\cov\left[\hat{x}(t)-x(t)\right] \preceq \gamma_c(A,C,Q,R) \cdot I,
			\end{align*}
			where $\gamma_e(A,C,Q,R)$ and $\gamma_c(A,C,Q,R)$ are scalars determined by system parameters $A,C,Q,R$ and independent of attack. $\Eb[\cdot],\cov[\cdot]$ are the expectation and the covariance with respect to the probability measure generated by the Gaussian noise.
		\end{problem}
		
		\begin{remark}
			In Problem \ref{pb:secure_est}, we only consider the estimation error at manipulated sampling times $t_k\in\Gamma^a$, because the estimation between sampling times $t\notin \Gamma^a$ is trivially provided by the following prediction 
			\begin{align*}
				\hat{x}(t)=\hat{x}(t_k)+ \exp({A(t-t_k)}) , t_k<t 
				<t_{k+1}, \ t_{k},t_{k+1}\in\Gamma^a
			\end{align*}
			and the estimation error are determined by $\hat{x}(t_k)$.
		\end{remark}
		
		\begin{remark}
			Even though we assume that $\Eb[x(0)]$ is known, the estimation of $x(t)$ is nontrivial when $A$ is unstable since the estimation error would be exponentially increasing if $x(t)$ is only predicted by $\exp(A t) \cdot \Eb[{x}(0)]$.
		\end{remark}
		
		With the help of the above assumptions and definitions, we are ready to deal with Problem~\ref{pb:secure_est} by designing a secure state estimation algorithm in the following section.		
		\section{Secure Estimation Design}
		In this section, we first introduce some preliminaries on local observable subspace decomposition. Secondly, the design of secure state estimation with the consideration of asynchronous measurement is presented. Finally, The analysis of resilient state estimation concludes the section.
		\subsection{Preliminaries on Local Observable Subspace Decomposition}\label{subsec:pre}
		Before introducing the local observable subspace decomposition, let us see a motivating example. 
		
		Consider the system \eqref{eq:system}-\eqref{eq:y_i_def} where $A$ is in the Jordan canonical form
		$$A=\begin{bmatrix}
			\lambda_1 &0&0\\
			0& \lambda_2 & 1\\
			0&0&\lambda_2
		\end{bmatrix}, \quad 
		C=\begin{bmatrix}
			1 &0&0\\
			0&1&0
		\end{bmatrix}.$$
		The system state is denoted as $x = 
		[x_1,~x_2,~x_3]^\top$.
		From the structure of the matrices $A$ and $C$,
		sensor 1 can only observe state $x_1$ with eigenvalue $\lambda_1$, while sensor 2 can observe the other states $x_2$ and $x_3$ with eigenvalue $\lambda_2$. This decoupling enables us to  study the following two subsystems separately:
		%
		\begin{align}
			\textbf{Subsystem 1: }&	\tilde{A}_1=\lambda_1,\; \tilde{C}_1=1.\\
			\textbf{Subsystem 2: }&\tilde{A}_2=\begin{bmatrix}
				\lambda_2 & 1\\
				0&\lambda_2
			\end{bmatrix}, \; \tilde{C}_2=\begin{bmatrix}
				1 &0
			\end{bmatrix}.
		\end{align}
		One can estimate the states of the subsystems 1 and 2 (denoted as $\tilde{x}_1 \in \Rb$ and $\tilde{x}_2 \in \Rb^2$) separately and then project back to the $3$-dimensional space. The projection is written as $H_1^\top \tilde{x}_1$ and $H_2^\top \tilde{x}_2$, where
		$$H_1=\begin{bmatrix}
			1&0&0
		\end{bmatrix}, \quad H_2=\begin{bmatrix}
			0&1&0\\ 0&0&1
		\end{bmatrix}.$$
		From the above illustrative example,
		let us show how to find the projection operation for a general linear system $(A,C)$ given $C_i$ as the row of the measurement matrix $C$ associated with sensor $i$.
		Define the observability matrix with respect to sensor~$i$ as
		\begin{equation}\label{eq:def_O}
			O_{i} \triangleq \begin{bmatrix}
				C_{i} ^\top&
				\left(C_{i} A\right){^\top}&\cdots&\left(C_{i} A^{n-1}\right){^\top}
			\end{bmatrix}^\top .
		\end{equation}
		Then, the local observable subspace of sensor~$i$ is defined
		{\small
			\begin{equation*}		\Ob_i\triangleq\rs(O_i)=\Span\left(C^\top_i,\left(C_{i} A\right){^\top},\cdots,\left(C_{i} A^{n-1}\right){^\top}\right).
			\end{equation*}
		}
        Denote the dimension of linear space $\Ob_i$ as $n_i\triangleq\dim(\Ob_i)$.
		The global observable space of the entire system is given by $\Ob\triangleq  \cup_{i\in\Ic} \Ob_i$ dut to the Jordan canonical form of $A$.
        Recalling Assumption~\ref{as:geo_mul} about the Jordan canonical form of matrix $A$, this assumption enables us to define the index set of states that can be observed by sensor $i$ as
		\begin{equation}\label{eq:def_Ec}
			\Qc_i\triangleq \{j\in \Jc \ |\ O_i \mathbf{e}_j\neq \mathbf{0} \}.
		\end{equation}
		Suppose the $n_i$ states can be observed from sensor $i$ are $\Qc_i=\{j_1,\cdots,j_{n_i}\}$, then the projection matrix from $\Ob$ to $\Ob_i$ is represented as follows:
		\begin{align}\label{eq:defH}
			H_i = 
			\begin{bmatrix}
				\mathbf{e}_{j_1}& \mathbf{e}_{j_2}& \cdots &\mathbf{e}_{j_{n_i}}
			\end{bmatrix}^\top\in\Rb^{n_i\times n}.
		\end{align}
		The following theorem characterizes the transformation between $\Ob$ and $\Ob_i$.
		\begin{theorem}\label{th:decomp_obs}
			If Assumption \ref{as:geo_mul} holds, for each sensor $i$, matrix $H_i$ represents the linear projection $\Ob\ra \Ob_i$ such that, for an arbitrary $x\in\Ob$, $H^\top_i H_i x=x$ if $x\in\Ob_i$, and $H^\top_i H_i x=\bm{0}$ if $x\in\Ob\setminus\Ob_i$.
		\end{theorem}
		\begin{proof}
			The proof directly follows our previous results \cite[Appendix A]{zishuo_timevary_full}.
		\end{proof}
		Theorem \ref{th:decomp_obs} provides an explicit formulation of the transformation from $\Ob$ to $\Ob_i$. 
		This plays a crucial role in our design of the decentralized observer that is resilient to spatio-temporal attacks. 		
		Define the state transition matrix from time $t$ to $t'$ as
		\begin{align}\label{eq:defAk}
			{\Lambda(t'-t)=\exp(A \cdot(t'-t) ).}
		\end{align}
    
		Define
		\begin{align}
			\tilde{A_i}(t)&\triangleq H_i \Lambda(t) H^\top_i \in\Cb^{n_i\times n_i},\label{eq:deftAk}\\
			\tilde{C_i}&\triangleq C_i H^\top_i \in\Cb^{1\times n_i}.
		\end{align} 
		The following properties of the transformation matrix $H_i$ are important to the design of local estimators.
		\begin{lemma}\label{lm:subspace}
			If Assumption \ref{as:geo_mul} holds, then the following properties hold  for all sensor $i\in\Ic,~t\in\Rb_{\geq0}$:
			\begin{align}
				H_i \Lambda(t)&=\tAi(t) H_i, \label{eq:HAH}\\
				C_i \Lambda(t)&=\tCi \tAi(t) H_i.\label{eq:CHH}
			\end{align}
			Moreover, if $(A,C)$ is observable, the pair $(\tAi(t),\tCi)$ is observable for all sensor $i\in\Ic,t\in\Rb_{\geq0}$.
		\end{lemma}
		\begin{proof}
			Notice that if $A$ is in Jordan canonical form, then $\exp(At),t>0$ is also in Jordan canonical form. Thus, the result directly follows our previous result \cite[Lemma 2]{zishuo_timevary_full}.
		\end{proof}
		Lemma~\ref{lm:subspace} affords us to design local state estimators of the decentralized observer when asynchronous non-periodic sensor measurements are considered in the following subsection.
		\subsection{Secure estimation with asynchronous measurements}\label{subsec:design}
		
		In the following, we propose a local estimator that maintains an estimate of $H_i x(t)$, i.e., the system state $x$ projected on local subspace $\Ob_i$. 
		For the convenience of denotation about the measurement process, we introduce the time-to-sensor index function $\psi$ as follows:
		\begin{align*}
			\psi(t)= \{i~|~(i,t,y_i(t))\in\Sc^a(t) \}.
		\end{align*}
		i.e., $\psi(t)$ denotes the sensor index of measurements with possibly manipulated time-stamp $t$. 	
		At each sampling time $t_k$, the dynamics of local estimate $\eta_i[k]$ corresponding to sensor $i$ is defined as
		\begin{align}\label{eq:eta}
			\eta_i[k]&=\tAi(t_{k}-t_{k-1})\eta_i[k-1]   \notag\\
			&+\mathbb{I}_{i\in\psi(t_k)} L_i[k] \left(y_i(t_{k})-\tCi \tAi(t_{k}-t_{k-1})\eta_i[k-1]\right),
		\end{align}
		where $\mathbb{I}_{i\in\psi(t_k)}$ is an indicator function which equals to 1 if $i\in \psi(t_k)$ or equals to 0 otherwise. 
		$\eta_i$ is initialized as $\eta_i[0]=H_i\Eb[x(0)]$. Here we assume that the expected initial state $\Eb[x(0)]$ is known. If it is unknown, we also have stable results (see Remark \ref{rm:Ex0}).
		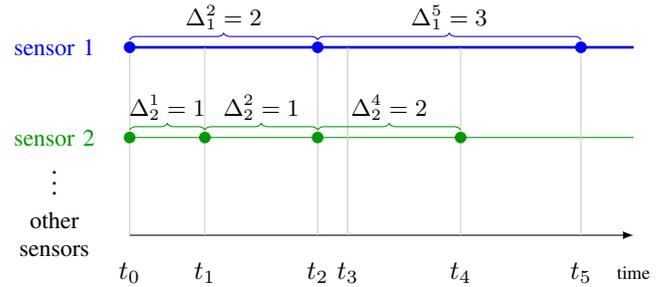
\begin{figure}[!ht]
			\centering
			\begin{tikzpicture}
\draw[blue,line width=1pt] (-4,4)--(2.7,4);
\node [blue] at (-5,4) {\small sensor 1};

\draw[green!60!black] (-4,2.8)--(2.7,2.8);
\node [green!60!black] at (-5,2.8) {\small sensor 2};

\node at (-5,2.3) {$\vdots$};
\node at (-5,1.7) {\small other};
\node at (-5,1.3) {\small sensors};

\draw[-latex][black] (-4,1.5)--(2.7,1.5);
\node at (2.7,1.0) {\scriptsize time};

\draw [black!20!white] (-4,1.5)--(-4,4);
\node at (-4,1.0) {$t_0$};

\draw [black!20!white] (-3,1.5)--(-3,4);
\node at (-3,1.0) {$t_1$};

\draw [black!20!white] (-1.5,1.5)--(-1.5,4);
\node at (-1.5,1.0) {$t_2$};

\draw [black!20!white] (-1.1,1.5)--(-1.1,4);
\node at (-1.1,1.0) {$t_3$};

\draw [black!20!white] (0.4,1.5)--(0.4,4);
\node at (0.4,1.0) {$t_4$};

\draw [black!20!white] (2,1.5)--(2,4);
\node at (2,1.0) {$t_5$};


\filldraw [blue] (-4,4) circle (2pt);

\filldraw [blue] (-1.5,4) circle (2pt);

\filldraw [blue] (2,4) circle (2pt);

\draw [blue, decorate, decoration = {brace}] (-4,4.1) --  (-1.5,4.1);
\node at (-2.75,4.4) {\small $\Delta_1^2=2$};

\draw [blue, decorate, decoration = {brace}]  (-1.5,4.1)--(2,4.1) ;
\node at (0.25,4.4) {\small $\Delta_1^5=3$};


\filldraw [green!60!black]  (-4,2.8) circle (2pt);

\filldraw [green!60!black]  (-3,2.8) circle (2pt);

\filldraw [green!60!black]  (-1.5,2.8) circle (2pt);

\filldraw [green!60!black]  (0.4,2.8) circle (2pt);

\draw [green!60!black, decorate, decoration = {brace}] (-4,2.9) --  (-3,2.9);
\node at (-3.5,3.2) {\small $\Delta_2^1=1$};

\draw [green!60!black, decorate, decoration = {brace}]  (-3,2.9)--(-1.5,2.9) ;
\node at (-2.25,3.2) {\small $\Delta_2^2=1$};

\draw [green!60!black, decorate, decoration = {brace}]  (-1.5,2.9)--(0.4,2.9) ;
\node at (-0.55,3.2) {\small $\Delta_2^4=2$};

\end{tikzpicture}\vspace{-10pt}
			\caption{The time interval notations and update illustration. The round dot denotes the corresponding sensor samples a measurement with the corresponding time-stamp. $\Delta_i^k$ is the number of time intervals since the last measurement sampled at sensor $i$ before time $t_k$.} \label{fig:time_axis}
		\end{figure}
		Define the stamp step index before sensor $i$ receives measurement with time-stamp $t_k$ as $t_{k-\Delta_i^k}$ (see Fig. \ref{fig:measurement_ava}) where
		\begin{align}
			\Delta_i^k\triangleq \min\{\Delta\in\Zb_{>0}| i\in\psi(t_{k-\Delta})\}, ~i\in\psi(t_k).
		\end{align}
		Notice that $\Delta_i^k$ is valid only when $i\in\psi(t_k)$.
		In the design of local estimators \eqref{eq:eta}, the estimator gain $L_i[k]$ should be designed such that
		\begin{align}\label{eq:I-LC_stable}
			\rho\left((I-L_i[k]\tCi)\tAi\left(t_k-t_{k-\Delta_i^k}\right) \right) \leq \bar{\rho}<1,
		\end{align}
		where $\bar{\rho}$ is a predefined design parameter to balance the weight between history estimation and new measurements.
		Since $\big( \tAi(t),\tCi \big)$ is observable from Lemma \ref{lm:subspace} and $\tAi(t)$ is non-singular for $t \in \Rb_{> 0}$, the system $\big(\tAi(t),\tCi\tAi(t) \big)$ is also observable. According to the Pole Assignment Theorem\cite{pole_assign}, there \textbf{always exists} $L_i[k]$ such that inequality in \eqref{eq:I-LC_stable} is satisfied. The following lemma ensures that such $L_i[k]$ not only exists but also has a uniformly bounded 2-norm.
		\begin{lemma}\label{lm:bounded_L}
			Suppose Assumptions \ref{as:geo_mul}-\ref{as:time_in_order} hold. There always exist a constant scalar $\bar{l}> 0$ determined by $A,C,\bar{\rho}$, and an estimator gain $L_i$ such that the following inequality holds for all $t \in (0,~T_{\max})$ with $T_{\max} > 0$ from Assumption \ref{as:sample_time}.
			\begin{align}
				\rho\left((I-L_i\tCi)\tAi\left(t\right) \right) \leq \bar{\rho}<1, ~\|L_i\|^2_2\leq \bar{l}.
			\end{align}
		\end{lemma}
		\begin{proof}
			See Appendix \ref{ap:proof_lm_bound_l}.
		\end{proof}

		The design of secure state estimation based on local observers is summarized in Algorithm~\ref{al:cen}. Due to the simple form of $H_i$, the fusion of all local estimates is done by taking the median:
			\begin{align}\label{eq:x^*=med}
				[\hat{x}[k]]_j={\med} \{[H_i^\top \eta_i[k]]_j, i\in\Fc_j \},
			\end{align}	 
			where $\Fc_j$ is designed as the index set of sensors that can observe state $j$.
			\begin{align}
				\Fc_j\triangleq \{i\in \Ic \ |\ O_i \mathbf{e}_j\neq \mathbf{0} \}.
			\end{align}
			In the following subsection, we will give the main theorem of this paper and provide its proof.

		\begin{algorithm}[h!t]
			\caption{Secure Estimation with asynchronous non-periodic measurements under attack}\label{al:cen}
			\begin{algorithmic}[1]
				\STATE {Receive a set of measurements with time-stamp $t_{k}$}
				\FOR {every sensor $i$}
				\IF {$i\in\psi(t_k)$}
				\STATE {Recall last time-stamp $t_{k-\Delta_i^k}$ when sensor $i$ has a measurement and calculate $\tAi\left(t_k-t_{k-\Delta_i^k}\right)$}
				\STATE{Obtain $L_i[k]$ with $\|L_i[k]\|^2_2\leq \bar{l}$ such that inequality \eqref{eq:I-LC_stable} is satisfied }
				\ENDIF
				\STATE Update $\eta_i[k]$ by \eqref{eq:eta}
				\ENDFOR
				\STATE Calculate each entry of $\hat{x}[k]$ by \eqref{eq:x^*=med}.				
			\end{algorithmic}
		\end{algorithm}
	\begin{remark}
		The $L_i[k]$ in line 5 always exists by Lemma \ref{lm:bounded_L}. One can design gain $L_i[k]$ such that the upper bound $\bar{l}$ can be purposely minimized, given $\bar{\rho}$.
    \begin{align*}
            L_i^\star[k] = &\arg \min_{L_i[k]} \|L_i[k]\|_2^2 \\
            \text{s.t.}&~  \|\succeq (I-L_i[k]\tCi)\tAi\big(t_k-t_{k-\Delta_i^k}\big)\|^2_2 \succeq \bar{\rho}^2 I
			\end{align*}
	\end{remark}
	
	\subsection{Main result}
	
	The following theorem is our main contribution of the paper. It claims that our proposed estimator is secure under spatio-temporal false data attack.
	Let us define the following constants:
	\begin{align}	\label{eq:defN}
		N_i=\int_{0}^{+\infty} s^i \cdot\frac{\rd}{\rd s} \Phi(s)^m\ \rd s ,\ i=1,2.
	\end{align}
	where $\Phi(s)$ is the cumulative density function of the standard Gaussian random variable. Define the initial estimation covariance spectral radius (not known to the estimator) as
	\begin{align}
		\sigma_0\triangleq\rho(\cov(\hat{x}(0)-x(0))).
	\end{align} 
	
	\begin{theorem}[secure estimation]\label{th:cen_lasso}
		Suppose Assumptions \ref{as:geo_mul}-\ref{as:time_in_order} are satisfied.
		The estimation error expectation is upper bounded by
		\begin{align}\label{eq:bounded_E}
			\left\|\Eb[\hat{x}[k]-x(t_k)]\right\|_{\infty} \leq  \left(\bar{\rho}^{2k}\sigma_0\bar{a}^2+\frac{(\bar{r}\bar{l}+\bar{q})\bar{a}^2}{1-\bar{\rho}^2}\right) N_1,
		\end{align}
		and the error covariance is bounded by
		\begin{align*}
			\rho\left(\cov[\hat{x}[k]-x(t_k)]\right)\leq\left(\bar{\rho}^{2k}\sigma_0\bar{a}^2+\frac{(\bar{r}\bar{l}+\bar{q})\bar{a}^2}{1-\bar{\rho}^2}\right)^2 N_2 .
		\end{align*}
$\bar{r},\bar{l},\bar{q},\bar{a}$ are scalar constants. The upper bound of measurement noise $\bar{r}$ is from Assumption \ref{as:R}. The upper bound of gain norm $\bar{l}$ is from Lemma \ref{lm:bounded_L}. The upper bound of process noise is $\bar{q}\triangleq \rho(Q)\cdot T_{\max}\cdot(\bar{l}\cdot\max_i{\|C_i\|_2}+1)^2$, and the upper bound of dynamic spectral radius is $\bar{a}\triangleq \sup_{0<t<T_{\max}}\rho(\exp(A\cdot t))$. 
	$\bar{\rho}$ is the design parameter in \eqref{eq:I-LC_stable}.
	\end{theorem}
	We will prove Theorem \ref{th:cen_lasso} in this subsection. The proof is based on the following two facts:
	\begin{enumerate}
		\item The local estimation error $\eta_i[k]-H_i x(t_k)$ is bounded for benign sensors.
		\item The estimation error of the median number is smaller than the maximum error of the benign sensors as long as redundancy is enough.
	\end{enumerate}
	
	We first prove the first statement by the following theorem. The second statement will be clear in the proof of Theorem~\ref{th:cen_lasso} given later.
	Define local estimation error as
	\begin{align}
		\epsilon_i[k]=\eta_i[k]-H_i x(t_k)
	\end{align}
The following theorem claims that the local estimation errors $\epsilon_i[k]$ of benign sensors are unbiased and stable.	
	\begin{theorem}\label{th:local_res}
		Consider benign sensors $i\in \Ic\setminus\Cc$. If Assumption \ref{as:geo_mul}-\ref{as:time_in_order} are satisfied and $L_i[k]$ satisfies the following inequalities:
		\begin{align*}
			\rho\left((I-L_i[k]\tCi)\tAi\left(t_k-t_{k-\Delta_i^k}\right) \right) \leq \bar{\rho},\ \|L_i[k]\|^2_2\leq \bar{l},
		\end{align*}  
		then the local residue $\epsilon_i[k]$ satisfies the following $\forall k\in\Zb_{\geq0}$
		\begin{align}
			\Eb(\epsilon_i[k]) =& 0 ,\\
			\rho(\cov(\epsilon_i[k]))\leq &\bar{\rho}^{2k}\sigma_0\bar{a}^2+\frac{(\bar{r}\bar{l}+\bar{q})\bar{a}^2}{1-\bar{\rho}^2}.
		\end{align}
		Thus, the local estimate of benign sensors are unbiased and has uniformly upper bounded covariance.
	\end{theorem}
	
	\begin{proof}
		
		Define $\tAi[k]\triangleq \tAi(t_k-t_{k-1})$ for notation simplicity.	
		For benign sensor $i \in \Ic \setminus \Cc$, if measurement from sensor $i$ is received with time-stamp $t_{k}$, from \eqref{eq:eta}, one has
		\begin{align*}
			\epsilon_i[k]=&\eta_i[k]-H_i x(t_{k})\\
			=&\tAi[k]\eta_i[k-1] \\
			&+ L_i[k] \left(y_i(t_{k})-\tCi \tAi[k]\eta_i[k-1]\right)\\
			& -H_i \Lambda(t_k-t_{k-1}) x(t_{k-1})-H_i w(t_{k-1},t_k)\\
			=&\left(\tAi[k]-L_i[k] \tCi\tAi[k]\right)\eta_i[k-1] \\
			+L_i[k]&\left[C_i \left(\Lambda(t_k-t_{k-1}) x(t_{k-1}) +w(t_{k-1},t_k) \right) +v_i(t_{k})\right]\\
			&-H_i  \Lambda(t_k-t_{k-1}) x(t_{k-1})-H_i w(t_{k-1},t_k)\\
			\overset{\rm (i)}{=}&\left(\tAi[k]-L_i[k] \tCi\tAi[k]\right)\eta_i[k-1] \\
			& - \tAi[k] H_i x(t_{k-1})+ L_i[k] \tCi \tAi[k]  H_i x(t_{k-1})  \\
			& +L_i[k]v_i(t_{k})+(L_i[k]C_i-H_i) w(t_{k-1},t_k) \\
			=&\left(\tAi[k]-L_i[k] \tCi\tAi[k]\right)\epsilon_i[k-1] +L_i[k]v_i(t_{k}) \\
			&+(L_i[k]C_i-H_i) w(t_{k-1},t_k),
		\end{align*}
		where equation (i) uses $H_i\Lambda(t)=\tAi(t)H_i$ and $C_i \Lambda(t)=\tCi \tAi(t)  H_i$ from Lemma \ref{lm:subspace}.

		In this scenario, the expectation and covariance dynamics are
		\begin{align*}
			\Eb[\epsilon_i&[k]]= \left(\tAi[k]-L_i[k] \tCi\tAi[k]\right) 	\Eb[\epsilon_i[k-1]]\\
			\cov(\epsilon_i&[k])=(I-L_i[k]\tCi)\tAi\left(t_k-t_{k-\Delta_i^k}\right)\times \\
			&\cov(\epsilon_i[k-\Delta_i^k])\tAi\left(t_k-t_{k-\Delta_i^k}\right)^\top (I-L_i[k]\tCi)^\top\\
			&+L_i[k]R_i(t_{k})L_i[k]^\top \\
            &+(L_i[k]C_i-H_i)Q (t_{k}-t_{k-1})(L_i[k]C_i-H_i)^\top .
		\end{align*}
		Based on Assumption \ref{as:R} and Lemma \ref{lm:bounded_L}, we have
		$$\rho\left(L_i[k]R_i(t_{k})L_i[k]^\top \right) \leq \bar{r} \rho\left(L_i[k]L_i[k]^\top\right) \leq \bar{r}\bar{l}.$$
		From Assumption \ref{as:sample_time} we have $t_k-t_{k-1}<T_{\max}$. Moreover, by the form of $H_i$ in \eqref{eq:defH} we know
\begin{align*}
    &\rho((L_i[k]C_i-H_i)Q (t_{k}-t_{k-1})(L_i[k]C_i-H_i)^\top)\\
    \leq &\rho(Q \cdot (t_k-t_{k-1})) \cdot (\|L_i[k]\|_2 \|C_i\|_2+\|H_i\|_2)^2 \leq \bar{q}.
\end{align*}		
  
		Suppose the variance of $x(0)$ is $\sigma_0$ (not known by the estimator). Then $\rho(\cov(\epsilon_i[0]))\leq \rho(\cov(\hat{x}(0)-x(0)))=\sigma_0$.
		Thus, if $i\in\psi(t_k)$, 
		the design condition \eqref{eq:I-LC_stable} implies that
		\begin{align}
			\rho(\cov(\epsilon_i[k]))\leq&  \bar{\rho}^{2k}\sigma_0+\sum_{t=0}^{k-1} \bar{\rho}^{2t}(\bar{r}\bar{l}+\bar{q})\notag \\
			\leq &\bar{\rho}^{2k}\sigma_0+\frac{\bar{r}\bar{l}+\bar{q}}{1-\bar{\rho}^2}. \label{eq:limsup_with_measurement} 
		\end{align}
		If no measurement triple from sensor $i$ is received with time-stamp $t_{k}$, the local residue satisfies
		\begin{align*}
			\Eb[\epsilon_i[k]]= & \left(\tAi[k]-L_i[k] \tCi\tAi[k]\right) 	\Eb[\epsilon_i[k-1]]\\
			\epsilon_i[k]=&\eta_i[k]-H_i x(t_{k})\\
			=&\tAi\left(t_k-t_{k-\Delta_i^k}\right) \eta_i[k-\Delta_i^k] \\
			&- H_i A \left(t_k-t_{k-\Delta_i^k}\right) x(t_{k-\Delta_i^k}) \\
			\overset{\rm (ii)}{=}& \tAi\left(t_k-t_{k-\Delta_i^k}\right) \epsilon_i[k-\Delta_i^k]
		\end{align*}
		where equation (ii) uses $H_i\Lambda(t)=\tAi(t)H_i$ from Lemma \ref{lm:subspace}. 
		Thus, if $i\notin\psi(t_k)$, since $i\in\psi(t_{k-\Delta_i^k})$ according to \eqref{eq:limsup_with_measurement}, we have
		\begin{align}\label{eq:limsup_without_measurement}
			\rho(\cov(\epsilon_i[k]))\leq &\bar{\rho}^{2k}\sigma_0\bar{a}^2+\frac{(\bar{r}\bar{l}+\bar{q})\bar{a}^2}{1-\bar{\rho}^2}.
		\end{align}
		Firstly, $\Eb[\epsilon_i[0]]= \Eb[\eta_i[k]-H_ix(0)]=H_i\Eb[x(0)]-H_i\Eb[x(0)]=0$ implies $\Eb[\epsilon_i[k]]=0$ for all $k\in\Zb_{\geq0}$.
	    Moreover,
		the results in \eqref{eq:limsup_with_measurement} and \eqref{eq:limsup_without_measurement} complete the proof.
	\end{proof}
	
	\begin{remark}\label{rm:Ex0}
		Since matrix $\tAi[k]-L_i[k] \tCi\tAi[k] $ is Schur stable, the equality
		$$	\Eb[\epsilon_i[k]]= \left(\tAi[k]-L_i[k] \tCi\tAi[k]\right) 	\Eb[\epsilon_i[k-1]]$$
		implies that even if $\Eb[x(0)]$ is unknown and $\eta_i[0]\neq H_i\Eb[x(0)] $, the expected local estimation error is converging to zero.
	\end{remark}
	
	
	We are now ready to present the proof of Theorem~\ref{th:cen_lasso}.
	\begin{proof}[\bf Proof of Theorem \ref{th:cen_lasso}]
		Let us define the following sequence order operator: $f_{i}\left(x_{l}, l\in\{1,\cdots,L\}\right)$ equals to the $i$-th smallest element in the set $\left\{x_{1}, \cdots, x_{L}\right\} .$ For even number $i$, we further define 
		$$f_{\frac{i+1}{2}} = \frac{1}{2} \left(f_{\frac{i}{2}} + f_{\frac{i}{2}+1}\right).$$ 
		Thus, $f_{(L+1)/2}\left(x_{l}, l\in\{1,\cdots,L\}\right)$ is the median number of set $\left\{x_{1}, \ldots, x_{L}\right\}$.
		From the definition of the sequence order operator $f_{i}$, we have an equivalent formulation to \eqref{eq:x^*=med} as follows:
		\begin{align*}
			[\hat{x}[k]-x(t_k)]_j = f_{(|\Fc_j|+1)/2}\left([H_i^\top \eta_i[k]-x(t_k)]_j, i\in\Fc_j\right) .
		\end{align*}
		Define the $\eta^o_{i}$ as the local estimate of sensor $i$ that is not manipulated by the attacker. 
		Since the system is $2p$-sparse observable and there are at most $p$ corrupted sensors, we have
		\begin{align*}
			\min_{i\in\Fc_j} \left\{ [\eta^o_i[k]-x(t_k)]_j \right\}\leq [\hat{x}[k]-x(t_k)]_j\\
			\leq \max_{i\in\Fc_j} \left\{ [\eta^o_i[k]-x(t_k)]_j\right\} .
		\end{align*}
		By employing Lemma \ref{lm:max_gaussian} in Appendix~\ref{app:lm_max_gaussian}, we have
		\begin{align*}
			-\sigma_{\max}[k] N_1 \cdot \bm{1}_n\leq\Eb[\hat{x}[k]-x(t_k)]\leq 	\sigma_{\max}[k] N_1 \cdot \bm{1}_n,
		\end{align*}
		where
		\begin{align}
			\sigma_{\max}[k]=\max_{i\in\Ic,j\in\{1,\cdots,n_i\}} \cov([\epsilon_i[k]]_j)\label{eq:covmax}
		\end{align}
		and \eqref{eq:bounded_E} is obtained.
		Due to 
		\begin{align*}
			\cov[\hat{x}[k]-x(t_k)]=\Eb[(\hat{x}[k]-x(t_k))^2]-\left(\Eb[\hat{x}[k]-x(t_k)]\right)^2,
		\end{align*}
		we have 
		\begin{align*}
			&\rho\left(\cov[\hat{x}[k]-x(t_k)]\right)\leq \\
			&\max\left\{ \Eb(\max\{\hat{x}[k]-x(t_k)\})^2, \Eb(\min\{\hat{x}[k]-x(t_k)\})^2 \right\}\\
			&\leq (\sigma_{\max}[k])^2 N_2
		\end{align*}
		Here the notations $\max\{\hat{x}[k]-x(t_k)\}$ and $\min\{\hat{x}[k]-x(t_k)\}$ represents the maximum and the minimum values of the vectors $\hat{x}[k]-x(t_k)$, respectively. The notation $\max\{a,b\}$ means the larger scalar between $a$ and $b$.
		Since $\sigma_{\max}[k] \leq \bar{\rho}^{2k}\sigma_0 \bar{a}^2+ \frac{(\bar{r}\bar{l}+\bar{q})\bar{a}^2}{1-\bar{\rho}^2}$ from Theorem \ref{th:local_res}, the results are obtained.	
	\end{proof}
	In the following section,
	we show numerical simulation results on IEEE 14-bus system to demonstrate the effectiveness of our proposed method.
	
	\section{Simulation}
	
\begin{figure}[!ht]
		\centering
		\includegraphics[width=0.4\textwidth]{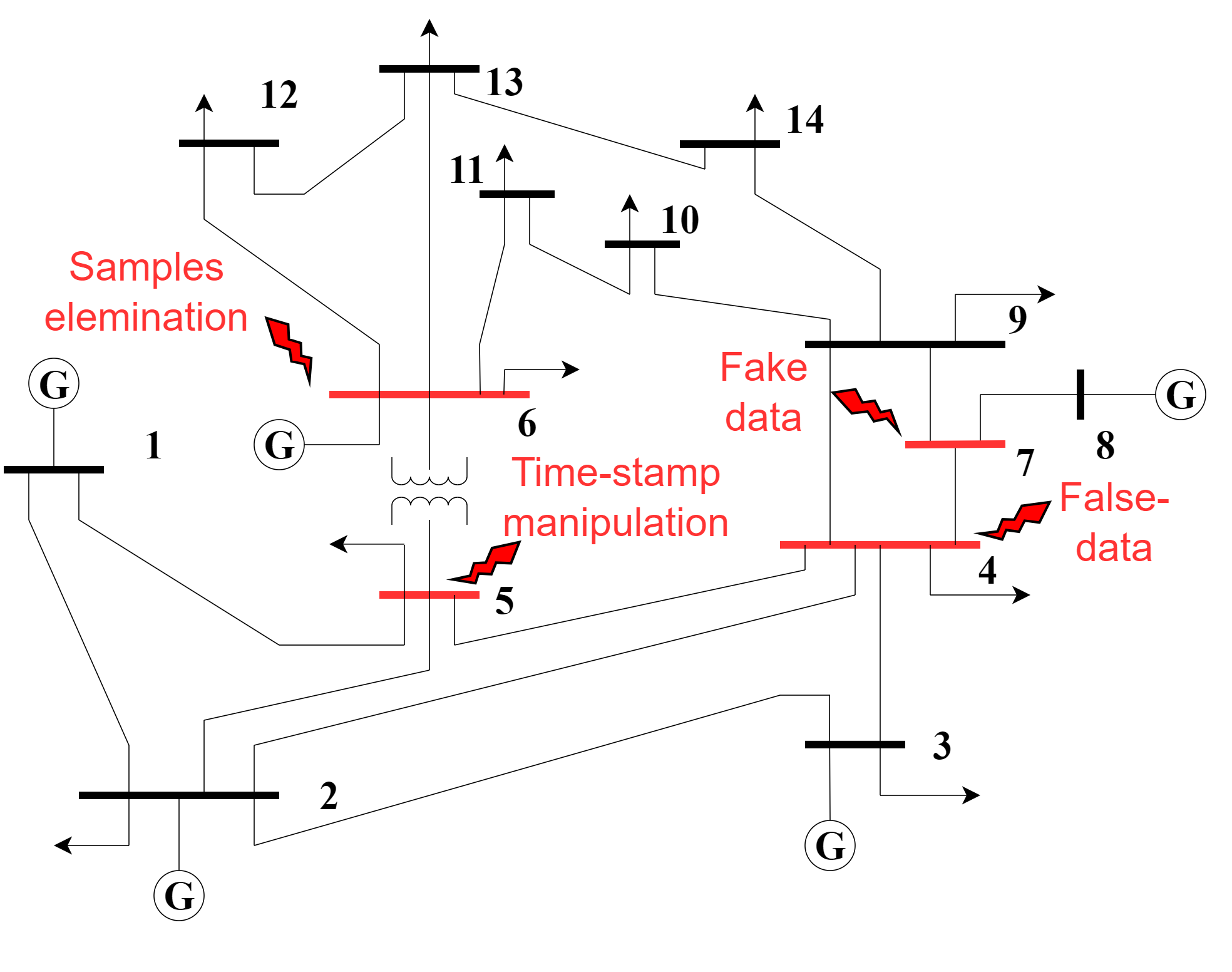}
		\caption{The IEEE 14-bus system consists of five generators on buses 1,2,3,6, and 8. Examples of spatio-temporal attacks: i) false-data attacks on bus 4; ii) time-stamp manipulation on bus 5; iii) samples elimination on bus 6; and iv) fake data on bus 7.}
		\label{fig:IEEE68}
	\end{figure}

\begin{figure}[!htb]
		\centering
		\input{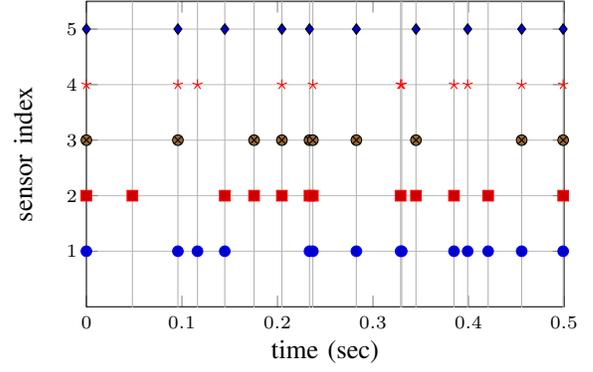}
		\caption{Sensor measurement availability of asynchronous non-periodic sampling. For conciseness, only show sensors 1-5 in the time interval of 0-0.5 seconds.} \label{fig:measurement_ava}
	\end{figure}

	We apply our proposed estimation scheme to the IEEE 14-bus system, which is a benchmark example extensively used in the literature \cite{George2010TPS,liu2020TSG} for illustrating the performance of secure state estimation algorithms. 
	The IEEE 14-bus system (see Fig. \ref{fig:IEEE68}) is composed of five generators. Let the generator bus index set be $\Vc_g=\{1,2,3,6,8\}$ and the load bus index set be $\Vc_l=\{2,3,4,5,6,9,10,11,12,13,14\}$. 
	We adopt the continuous-time system dynamics as in the following equations \cite{AllenPowerbook2013}:
	\begin{align*}
		\dot{\theta}_{i}(t) &=\omega_{i}(t), \\
		\dot{\omega}_{i}(t) &=-\frac{1}{m_{i}}\left[D_{i} \omega_{i}(t)+\sum_{j \in \mathcal{N}_{i}} P_{t i e}^{i j}(t)- P_{i}(t)+w_{i}(t)\right],
	\end{align*}
	where $\theta_{i}(t)$ and $\omega_{i}(t)$ are the phase angle and angular frequency on bus $i$, respectively, $m_i$ is the angular momentum of $i$, and $w_i$ is the process disturbance.
	The parameter $D_i$ is the load change sensitivity w.r.t. the frequency  \cite[Section 10.3]{AllenPowerbook2013}.
	The power flow between neighboring buses $i$ and $j$ is given by $P_{\rm tie}^{i j}(t)=-P_{t i e}^{j i}(t)=t_{i j}\left(\theta_{i}(t)-\theta_{j}(t)\right)$, where $t_{ij}$ is the inverse of resistance between bus $i$ and $j$. 
	The power $P_{i}(t)$ denotes the difference between the mechanical power and power demand at bus $i$, which is known by the system operator. 
	Every bus is equipped with 3 sensors: one electric power sensor, one phase sensor, and one angular velocity sensor.
	The measurements are sampled non-periodically with sampling intervals uniformly distributed in $[0.001,0.05]$, and each sensor has a probability of 0.6 of successful sampling at each time stamp (see Fig. \ref{fig:measurement_ava}). The measurement noise covariance is $Q=0.001\cdot I, R=0.01\cdot I$.

\begin{figure}[!htb]
		\centering
		\input{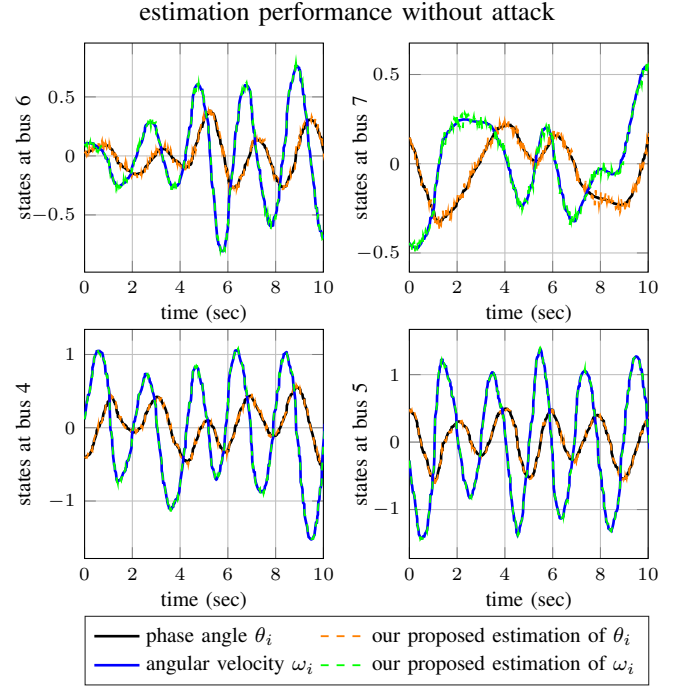}
		\caption{Estimation of states without attack.} \label{fig:IEEE14_no}
	\end{figure}

	Figure \ref{fig:IEEE14_no} shows the estimation performance without an adversary, the estimation holds stable but relatively large due to the conservativeness of median fusion algorithm.
	
\begin{figure}[!htb]
		\centering
		\input{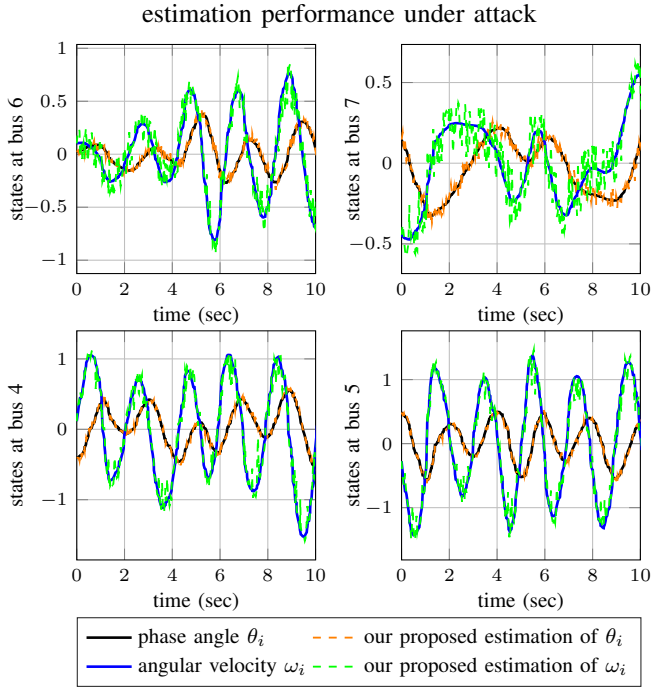}
		\caption{Estimation performance under attack.} \label{fig:IEEE14_attack}
	\end{figure}

	The attack is launched on angular velocity sensors of buses $4,5,6$, and $7$. At bus $4$, random false data is injected into its measurements. At bus $5$, the time-stamp is randomly shifted (and thus the order of samples is changed). At bus $6$, the samples are eliminated with probability 0.5. At bus $7$, new fake data with random measurements and time-stamps are generated.
	Fig. \ref{fig:IEEE14_attack} demonstrates the estimation performance on buses 5-7 under the attacks mentioned before. The estimation error is slightly larger than the case without attack but still remains stable despite the various false data attacks.

	 \appendix
	
	 \subsection{Proof of Lemma \ref{lm:bounded_L}}\label{ap:proof_lm_bound_l}
	 \begin{proof}
	 	Suppose the eigenvalues of $H_iAH_i^\top$ as $\lambda_1,\cdots,\lambda_{n_i}$. Then the eigenvalues of $\tAi(t)$ are $e^{\lambda t_1},\cdots,e^{\lambda t_{n_i}}$. Thus the characteristic polynomial of $\tAi(t)$ is $p_{i,t}(s)=\prod_{j=1}^{n_i} (s-e^{\lambda t_{j}})$. Denote the coefficients of the $n_i$-degree polynomial $p_{i,t}(s)$ as $\{c_{i,n_i}(t),c_{i,n_i-1}(t),\cdots,c_{i,1}(t)\}$. According to the form of $p_{i,t}(s)$, set $\bigcup_{0<t<T_{\max}}\{c_{i,n_i}(t),c_{i,n_i-1}(t),\cdots,c_{i,1}(t)\} $ is compact. 
		
	 	Since $(\tAi(t), \tCi\tAi(t))$ is observable for all $t>0$, there exists a non-singular transformation $V_i(t)$ such that $(V_i^{-1}(t)\tAi(t) V_i(t), \tCi \tAi(t) V_i(t))=(\check{A}_i(t), \check{C}_i(t))$ is the observable canonical form. Further, the pole assignment theorem \cite{pole_assign} gives us the following observer gain
	 	\begin{align}
	 		L_i[k]=V_i(t_k-\Delta_i^k)\cdot \begin{bmatrix}
	 			r^i_{n_i}(\bar{\rho}) - c_{i,n_i}(t_k-\Delta_i^k)\\
	 			\vdots \\
	 			r^i_{1 }(\bar{\rho}) - c_{i,1}(t_k-\Delta_i^k)
	 		\end{bmatrix}, \label{lempf:L_i}
	 	\end{align} 
	 	where $\{r^i_{n_i}(\bar{\rho}),\cdots,r^i_{1}(\bar{\rho}) \}$ are the coefficients of characteristic polynomials corresponding to assigned eigenvalues determined by $\bar{\rho}$.
		
	 	Since the observability matrix of system $\left(\tAi(t),\tCi\tAi(t)\right)$ is bounded away from singularity, we know $\bigcup_{0<t<T_{\max}} \{V_i(t)\}$ belongs to a compact set. Moreover, recall that $\bigcup_{0<t<T_{\max}}\{c_{i,n_i}(t),c_{i,n_i-1}(t),\cdots,c_{i,1}(t)\} $ belongs to a compact set, the observer gain $L_i[k]$ in \eqref{lempf:L_i} has uniformly upper bounded 2-norm, completing the proof. 
		
	 \end{proof}
	 \subsection{Bounds on maximum and minimum of Gaussian random variables}
	 \label{app:lm_max_gaussian}
	 \begin{lemma}\label{lm:max_gaussian}
	 	Suppose $X_i,i=1,2,\cdots,m$ are independent Gaussian random variables with expectation $\Eb[X_i]=0$ and standard deviation $\sigma_{i}$. 
	 	Define $N_1,N_2$ as in \eqref{eq:defN}. Then, one has the following results
	 	\begin{align*}
	 		\Eb\left[\max_{i=1,2,\cdots,m} X_i\right]&\leq \left(\max_i \sigma_{i}\right)\cdot N_1 , 	\\
	 		\Eb\left[\min_{i=1,2,\cdots,m} X_i\right]&\geq -\left(\max_i \sigma_{i}\right)\cdot N_1. \\
	 		\Eb\left[\left(\max_{i=1,2,\cdots,m} X_i\right)^2\right]&\leq \left(\max_i \sigma_{i}\right)^2\cdot N_2 , 	\\
	 		\Eb\left[\left(\min_{i=1,2,\cdots,m} X_i\right)^2\right]&\leq \left(\max_i \sigma_{i}\right)^2\cdot N_2.
	 	\end{align*}
	 \end{lemma}
	 \begin{proof}
	 	Define constant 
	 	\begin{align*}
	 		N_1^-=&\int_{-\infty}^0 s \cdot\frac{\rd}{\rd s} \Phi(s)^m\ \rd s<0,
	 	\end{align*}
	 	where $\Phi(\cdot)$ as the CDF of standard normal distribution.
	 	\begin{align*}
	 		\mathbb{P}\left(\max_i X_i < s\right) = \prod_{i=1}^m \Phi\left(\frac{s}{\sigma_i}\right)\leq \begin{cases}
	 			\left[\Phi\left(\frac{s}{\sigma_{\max}}\right)\right]^m \text{ if } s\leq0\\
	 			\left[\Phi\left(\frac{s}{\sigma_{\min}}\right)\right]^m \text{ if } s>0
	 		\end{cases} 
	 	\end{align*}
	 	where $\sigma_{\min}=\min_{i} \sigma_{i}, \sigma_{\max}=\max_{i} \sigma_{i}$.
		
	 	Thus,
	 	\begin{align*}
	 		&\mathbb{E}\left( \max_i X_i \right)= \int_{-\infty}^{+\infty} s \cdot {\rd} \left(\prod_{i=1}^m \Phi\left(\frac{s}{\sigma_i}\right) \right)\\
	 		\leq &\int_{-\infty}^{0} s \cdot \frac{\rd}{\rd s} \left[\Phi\left(\frac{s}{\sigma_{\max}}\right)\right]^m {\rd s}\\
	 		&+\int_{0}^{+\infty} s \cdot \frac{\rd}{\rd s} \left[\Phi\left(\frac{s}{\sigma_{\min}}\right)\right]^m {\rd s}\\
	 		=&\sigma_{\max}\int_{-\infty}^{0} s \cdot \frac{\rd}{\rd s} \left[\Phi\left(s\right)\right]^m {\rd s}+\sigma_{\min}\int_{0}^{+\infty} s \cdot \frac{\rd}{\rd s} \left[\Phi\left(s\right)\right]^m {\rd s}\\
	 		=&\sigma_{\max}\cdot N_1^{-}+\sigma_{\min}\cdot N_1 \leq \sigma_{\min}\cdot N_1\leq \sigma_{\max}\cdot N_1.
	 	\end{align*}
	 	Similarly,
	 	\begin{align*}
	 		&\mathbb{P}\left(\min_i X_i < s\right) = 1-\mathbb{P}\left(\min_i X_i > s\right)\\
	 		=&1-\prod_{i=1}^m \left[1- \Phi\left(\frac{s}{\sigma_i}\right)\right]\geq
	 		\begin{cases}
	 			1-\left[1- \Phi\left(\frac{s}{\sigma_{\min}}\right)\right]^m \text{if }s\leq 0\\
	 			1-\left[1- \Phi\left(\frac{s}{\sigma_{\max}}\right)\right]^m \text{if }s>0
	 		\end{cases} 
	 	\end{align*}
	 	Notice that we have the following fact:
	 	\begin{align}
	 		&\int_{-\infty}^{0} s \cdot \frac{\rd}{\rd s} \left(1-\left[1- \Phi\left({s}\right)\right]^m\right) {\rd s}\notag \\
	 		=&\int_{+\infty}^0 -s \cdot {\rd} \left(1-\left[1- \Phi\left(-s\right)\right]^m\right)\notag \\
	 		=&-\int_{0}^{+\infty} s \cdot \frac{\rd}{\rd s} \left[\Phi\left(s\right)\right]^m {\rd s},		\label{eq:phis+pshi-s}
	 	\end{align}
	 	where the equality \eqref{eq:phis+pshi-s} comes from $\Phi(s)+\Phi(-s)=1$. Similarly we have $\int_{0}^{+\infty} s \cdot \frac{\rd}{\rd s} \left(1-\left[1- \Phi\left({s}\right)\right]^m\right) {\rd s}=-\int_{-\infty}^{0} s \cdot \frac{\rd}{\rd s} \left[\Phi\left(s\right)\right]^m {\rd s}.$
	 	Thus,
	 	\begin{align*}
	 		&\mathbb{E}\left( \min_i X_i \right)\\
	 		\geq &\int_{-\infty}^{0} s \cdot \frac{\rd}{\rd s} \left(1-\left[1- \Phi\left(\frac{s}{\sigma_{\min}}\right)\right]^m\right) {\rd s}\\
	 		&+\int_{0}^{+\infty} s \cdot \frac{\rd}{\rd s} \left(1-\left[1- \Phi\left(\frac{s}{\sigma_{\max}}\right)\right]^m\right) {\rd s}\\
	 		=&-\sigma_{\min}\cdot N_1-\sigma_{\max}\cdot N_1^-\\
	 		\geq& -\sigma_{\min}\cdot N_1\geq -\sigma_{\max}\cdot N_1.
	 	\end{align*}
		
	 	For the second order moment and $s\geq0$, we have
	 	{\scriptsize
	 		\begin{align*}
	 			\mathbb{P}\left(\left(\max_i X_i\right)^2 < s\right) \leq& \mathbb{P}\left( \left(\max_i X_i\right) < \sqrt{s}\right) \leq\left[\Phi\left(\frac{\sqrt{s}}{\sigma_{\min}}\right)\right]^m , \\
	 			\mathbb{P}\left(\left(\min_i X_i\right)^2 < s\right) \leq& 1-\mathbb{P}\left( \left(\min_i X_i\right) > \sqrt{s}\right)\\ \leq&1-\left[1-\Phi\left(\frac{\sqrt{s}}{\sigma_{\min}}\right)\right]^m , 
	 	\end{align*}}
	 	Similar to previous deduction, we have 
	 	\begin{align*}
	 		&\mathbb{E}\left(\left( \max_i X_i \right)^2\right)\leq\int_{0}^{+\infty} s \cdot \frac{\rd}{\rd s} \left[\Phi\left(\frac{\sqrt{s}}{\sigma_{\min}}\right)\right]^m {\rd s}\\
	 		=&\sigma_{\min}^2\int_{0}^{+\infty} s^2 \cdot \frac{\rd}{\rd s} \left[\Phi\left(s\right)\right]^m {\rd s}\\
	 		=&\sigma_{\min}^2\cdot N_2 \leq \sigma_{\max}^2\cdot N_2.
	 	\end{align*}
	 	and $\mathbb{E}\left(\left( \min_i X_i \right)^2\right)\leq \sigma_{\max}^2\cdot N_2$.
	 \end{proof}

	\bibliographystyle{IEEEtran}
	\bibliography{ref_cdc2023_timestamp}
	
\end{document}